\documentclass[dvips,, 12pt]{article}
\usepackage{latexsym, epsfig, amsmath, graphicx, amssymb}
\usepackage{mathrsfs, amsfonts, graphics}
\usepackage{graphicx}

\begin{document}

\begin{center}
\textbf{\Large Charged Rotating BTZ Black Hole and Thermodynamic
Behavior of Field Equations at its Horizon}\\[1.2cm]
M. Akbar$^{1}$ and Azad A. Siddiqui$^{2}$\\[3ex]
$^{1}$Centre for Advanced Mathematics and Physics, National
University of Sciences and Technology, Campus of EME College,
Peshawar Road, Rawalpindi,
Pakistan \\[0pt]
E-mail: ak64bar@yahoo.com \\[0pt]
$^{2}$Department of Basic Science and Humanities, EME College,
National University of Science and Technology, Peshawar Road,
Rawalpindi, Pakistan
\\[0pt]
E-mail: azad@ceme.edu.pk \\[0pt]

\bigskip

\bigskip

\bigskip

\textbf{Abstract}
\end{center}

\begin{quotation}
In this paper, we study different cases of the charged rotating
BTZ black hole with reference to their horizons. For the existence
of these cases conditions on mass, charge and angular momentum of
the black hole are obtained. It is also shown that the Einstein
field equations for the charged rotating BTZ black hole at the
horizon can be expressed as first law of thermodynamics,
$dE=TdS+\Omega dJ+\Phi dq+P_{r}dA$.
\end{quotation}

PACS Numbers: 04.70.Dy, 97.60.Lf

\newpage

\section{Introduction}

The relationship between the Einstein field equations and the
first law of black hole thermodynamics was first presented by
Jacobson \cite{jac}. Padmanabhan \cite {a6} made the key
development by establishing a general formalism for understanding
the thermodynamics of horizons in spherically symmetric spacetimes
and showed that it is possible to write the Einstein field
equations in the form of first law of thermodynamics arising from
the virtual displacement of the horizon normal to itself as
\begin{equation}
TdS=dE+PdV.  \label{1}
\end{equation}
Paranjape, Sarkar and Padmanabhan \cite{PSP} studied this approach
for a more general Lanczos-Lovelock theories of gravity and found
that it is possible to interpret the field equations near horizon
for spherically symmetric spacetimes as a thermodynamic identity
$\left( \ref{1}\right)$. The Lanczos-Lovelock gravity contains
higher derivative terms which can be thought as quantum
corrections to the Einstein gravity, therefore, it may be expected
that the thermodynamic interpretation of the field equations holds
even if one includes possible quantum corrections to the
Einstein-Hilbert action. Kothawala, Sarkar and Padmanabhan
\cite{ksp} further studied this approach for stationary
axis-symmetry horizons and time dependent evolving horizons and
found, in both cases, that the field equations near horizon can be
expressed as a thermal identity. In references \cite{a10, a15} it
has been shown that the Einstein field equations at apparent
horizon of the Friedmann universe can also be expressed as a
thermal identity.

These thermodynamic interpretations of gravitational dynamics at
horizons need further investigation for understanding it at a
deeper level \cite{a6, a5}. In recent years, (2+1)-dimensional BTZ
(Banados-Teitelboim-Zanelli) black holes have drawn a lot of
attention as simplified models for exploring conceptual issues
relating to the black hole thermodynamics (see, e.g., \cite{btz,
btz1}).

Recently, the thermodynamic interpretation of field equations for
static as well as non-static BTZ black hole near horizon is
presented in \cite{akb}. In this paper we have extended the
thermodynamic study for charged rotating BTZ (CR-BTZ) black hole.
After describing the CR-BTZ black hole in the following section,
which has also been discussed in reference \cite{JZ}, we will
discuss the thermodynamic interpretation of the Einstein field
equations in section-III and finally the conclusion is given in
the last section.

\section{Charged Rotating BTZ Black Hole}

\bigskip Consider Einstein field equations for (2+1)-dimensional spacetime
with negative cosmological constant $\Lambda =-l^{-2}$
\begin{equation}
G_{ab}+\Lambda g_{ab}=\pi T_{ab},\text{ \ \ \ \ \ \ \ \ }\left(
a,b=0,1,2\right)  \label{2}
\end{equation}
where $G_{ab}$ and $T_{ab}$\ are the Einstein and stress-energy
tensors respectively. For the electro-vacuum stress-energy tensor,
the standard CR-BTZ solution of the above field equations $\left(
\ref{2}\right) $ is given by the following metric \cite{a17, mann}
\begin{equation}
ds^{2}=-f(r)dt^{2}+\frac{dr^{2}}{f(r)}+r^{2}(d\phi
-\frac{J}{2r^{2}}dt)^{2}, \label{3}
\end{equation}
where the lapse function $f(r)$, in the units such that the
(2+1)-dim gravitational constant $G_{3}=\frac{1}{8}$, is given by
\begin{equation}
f(r)=-M+\frac{r^{2}}{\ell ^{2}}+\frac{J^{2}}{4r^{2}}-\frac{\pi
}{2} Q^{2}lnr.  \label{4}
\end{equation}
Here $M$ and $J$ are the integration constants which respectively
correspond to the mass and angular momentum, and $Q$ is the charge
carried by the black hole.

Horizons of the CR-BTZ metric $\left( \ref{3}\right)$ are roots of
the lapse function. Depending on these roots there are three cases
of the CR-BTZ black hole, which we name (analogous to the Reissner
Nordstrom or Kerr-Newman spacetimes) as:

(i) Usual CR-BTZ black hole when two distinct real roots exist

(ii) Extreme CR-BTZ black hole in case of two repeated real roots

(iii) Naked CR-BTZ singularity when no real root exists.

\bigskip It is hard to find exact expressions for the roots of the lapse
function, however we analyze the function qualitatively in order
to obtain conditions on $M$, $J$ and $Q$ for the existence of
above mentioned three cases as follows:

Extremal points of the function are determined using
$\frac{df}{dr}=0$, which gives $r^{2}=\ell ^{2}(\pi Q^{2}\pm
\sqrt{\pi ^{2}Q^{4}+16J^{2}/\ell ^{2}})/8$. Excluding the negative
sign as it gives $r^{2}<0$, we have a unique extremal point
$r=r_{\text{min}}=\ell \sqrt{(\pi Q^{2}+\sqrt{\pi
^{2}Q^{4}+16J^{2}/\ell ^{2}})/8}$ (here again the negative root is
excluded). This extremal point is a minima as
$\frac{d^{2}f}{dr^{2}}>0$ and the minimum value of the function is

\begin{eqnarray}
f(r_{min}) &=&-M+\frac{\pi Q^{2}+\sqrt{\pi
^{2}Q^{4}+\frac{16J^{2}}{\ell ^{2} }}}{8}+\frac{2J^{2}}{\ell
^{2}(\pi Q^{2}+\sqrt{\pi ^{2}Q^{4}+\frac{
16J^{2}}{ \ell ^{2}}})}  \notag \\
&&-\frac{\pi }{4}Q^{2}ln\left[ \frac{\ell ^{2}}{8}\left( \pi
Q^{2}+\sqrt{\pi ^{2}Q^{4}+\frac{16J^{2}}{\ell ^{2}}}\right)
\right] .  \label{5}
\end{eqnarray}
Further, the value of $f(r)\rightarrow \infty $ both as
$r\rightarrow 0$ or $ r\rightarrow \infty $. Now consider the
cases

(a) $f(r_{min})<0$,~~~ (b) $f(r_{min})=0$,~~~ (c) $f(r_{min})>0$.
\newline
\newline
\textbf{Case a:} In this case $f(r)$ intersects the $r$-axis at
two points and hence we have two distinct real roots at $r=r_{\pm
}$ (see Figure 1) and $f(r_{min})<0$ implies

\begin{figure}[htb]
  \begin{center}
  \includegraphics[width=.5\textwidth]{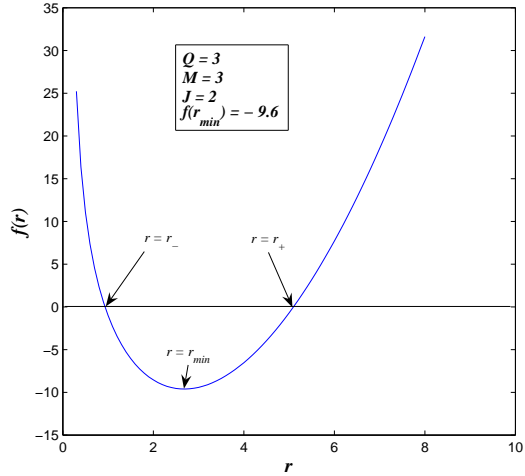}\\
  \caption{\emph{Lapse function $f(r)$ for $Q=3$, $M=3$ and $J=2$ is shown which
  cuts the r-axis at two distinct points $r=r_{\pm}$. Here $f(r_{min})$
  is negative. $\ell ^{2}=1$ in all figures.}}
\end{center}
\end{figure}

\begin{equation}
M>\frac{\pi Q^{2}+\sqrt{\pi ^{2}Q^{4}+\frac{16J^{2}}{\ell
^{2}}}}{8}+\frac{ 2J^{2}}{\ell ^{2}(\pi Q^{2}+\sqrt{\pi
^{2}Q^{4}+\frac{16J^{2}}{\ell ^{2}}})}- \frac{\pi
}{4}Q^{2}ln\left[ \frac{\ell ^{2}}{8}\left( \pi Q^{2}+\sqrt{\pi
^{2}Q^{4}+\frac{16J^{2}}{\ell ^{2}}}\right) \right] ,  \label{6}
\end{equation}
Notice that the above inequality $\left( \ref{6}\right) $ is the
general condition for the existence of two horizons for the CR-BTZ
black hole. Its particular cases (i) for $Q=0$ and $J\neq 0$
reduces to $M>J/\ell$ which is a known condition for the
non-static BTZ black hole \cite {akb} and (ii) for $J=0$ and
$Q\neq 0$ it reduces to $M>\frac{\pi Q^{2}}{4} \left[ 1-\ln \left(
\frac{\pi \ell ^{2}Q^{2}}{4}\right) \right] $ which is also known
for the static charged BTZ black hole \cite{JZ}.
\newline
\textbf{Case b:} This corresponds to the extreme CR-BTZ black
hole. In this case we have two real repeated roots at
$r=r_{-}=r_{+}$ (see Figure 2) and $M$ is equal to the right hand
side of the inequality $\left( \ref{6}\right)$. The particular
cases (i) $Q=0$ and $J\neq 0$ implies $M=J/\ell$ and (ii) $J=0$
and $Q\neq 0$ implies $M=\frac{\pi Q^{2}}{4} \left[ 1-\ln \left(
\frac{\pi \ell ^{2}Q^{2}}{4}\right) \right]$.

\begin{figure}[htb]
  \begin{center}
  \includegraphics[width=.5\textwidth]{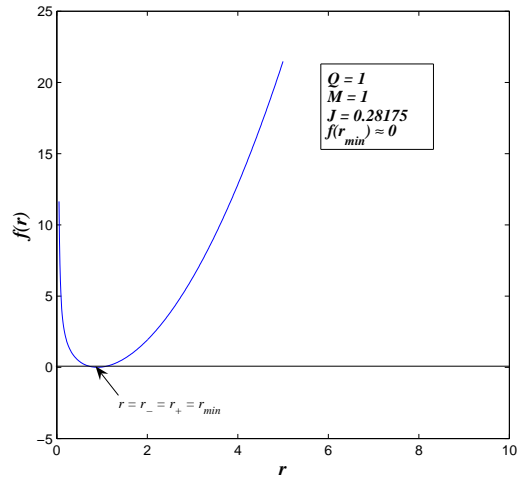}\\
  \caption{\emph{Lapse function $f(r)$ for $Q=1$, $M=1$ and $J=0.28175$ is
  shown which touches the r-axis at one point $r=r_{-}=r_{+}=r_{\min}$.
  Here $f(r_{min})\approx 0$.}}
\end{center}
\end{figure}

\textbf{Case c: }This corresponds to the Naked CR-BTZ singularity.
In this case there are no real roots (see Figure 3) and $M$ is
less than the right hand side of the inequality $\left(
\ref{6}\right)$.

\begin{figure}[htb]
  \begin{center}
  \includegraphics[width=.5\textwidth]{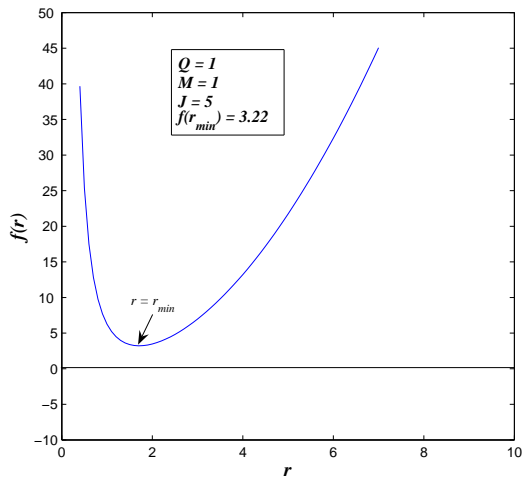}\\
  \caption{\emph{Lapse function $f(r)$ for $Q=1$, $M=1$ and $J=5$ is
  shown which does not intersect the r-axis. Here $f(r_{min})$ is positive.}}
\end{center}
\end{figure}

\subsection{Heat Capacity of the CR-BTZ Black Hole }

\bigskip The heat capacity at the horizon of the CR-BTZ black hole, using
relation $C_{J,Q}=(\partial M/\partial T)_{J,Q}$, is obtained as
\begin{equation}
C_{J,Q}=4\pi r_{+}\left( \frac{4r_{+}^{4}-J^{2}\ell ^{2}-\pi
Q^{2}\ell ^{2}r_{+}^{2}}{4r_{+}^{4}+3J^{2}\ell ^{2}+\pi Q^{2}\ell
^{2}r_{+}^{2}} \right) .  \label{6a}
\end{equation}
The heat capacity is positive if the term in the numerator of
equation  $ \left( \ref{6a}\right) $ is positive i.e.
\begin{equation}
4r_{+}^{4}-J^{2}\ell ^{2}-\pi Q^{2}\ell ^{2}r_{+}^{2}>0.
\label{6b}
\end{equation}
Completing square on left hand side of the inequality $\left(
\ref{6b}\right) $ and simplifying we get
\begin{equation}
r_{+}>r_{min}.  \label{6c}
\end{equation}
Therefore $C_{J,Q}>0$ if $r_{+}>r_{min}$, which holds for the
usual CR-BTZ black hole case showing its stability. For the
extreme CR-BTZ black hole case $r_{+}=r_{min}$ and $C_{J,Q}=0$.
Notice that for $Q=0$ both, the expression for the heat capacity
given by equation $\left( \ref{6a}\right) $ and condition  $\left(
\ref{6c}\right)$ reduce to the expression and condition obtained
by Cai et al \cite{Cai} for the non-static BTZ black hole.

\section{Thermodynamic Interpretation of the Einstein Field Equations}

In this section we discuss thermodynamic properties of the CR-BTZ
black hole at the event horizon located at $r=r_{+}$. This horizon
is associated with temperature $T$, entropy $S$, angular momentum
$J$ and electric potential $ \Phi$ as

\begin{equation}
T=\frac{1}{4\pi }\frac{df}{dr}|_{r=r_{+}},S=4\pi
r_{+},J=2r_{+}^{2}\Omega =2r_{+}^{2}\frac{\partial M}{\partial
J}|_{r=r_{+}},\Phi =\frac{\partial M}{
\partial Q}|_{r=r_{+}}=-\pi Qlnr_{+}.  \label{7}
\end{equation}
where $\Omega $ is the angular velocity. The mass $M$ at $r=r_{+}$
is given by
\begin{equation}
M=\frac{r_{+}^{2}}{\ell ^{2}}+\Omega ^{2}r_{+}^{2}+\frac{\Phi
}{2}Q. \label{8}
\end{equation}
Note that the thermodynamic quantities $T$,  $S$, $J$, $Q$ and $M$
obey the first law of thermodynamics $dM=TdS+$ $\Omega dJ+\Phi
dQ$.

Taking differential of equation $\left( \ref{8}\right)$ we have
\begin{equation}
dM=\frac{2r_{+}}{\ell ^{2}}dr_{+}+2\Omega
^{2}r_{+}dr_{+}+\frac{\Phi }{2}dQ. \label{9}
\end{equation}

The (1, 1)-component of the field equation $\left( \ref{2}\right)
$, for the metric $\left( \ref{3}\right) $, for a general $f(r)$
can be written in the form
\begin{equation}
\frac{J^{2}+2r^{3}f^{\prime }(r)}{4r^{4}}-\frac{1}{\ell ^{2}}=\pi
T_{1}^{1}, \label{10}
\end{equation}
where $T_{1}^{1}$ is the (1, 1)-component of the stress-energy
tensor which corresponds to the radial pressure
($P_{r}=-T_{1}^{1}$) of the source. Equation $\left(
\ref{10}\right) $ when evaluated at $r=r_{+}$ can be written as
\begin{equation}
\frac{J^{2}}{4r_{+}^{3}}+\frac{1}{2}f^{\prime
}(r_{+})-\frac{r_{+}}{\ell ^{2} }=-P_{r}(\pi r_{+}),  \label{11}
\end{equation}
The above equation $\left( \ref{11}\right) $ describes the
dynamics of the spacetime near horizon. In order to determine a
possible thermodynamic interpretation of this equation near
horizon, consider a virtual displacement $dr_{+}$ of the horizon
and multiply it on both sides of the equation. Then the resulting
equation can be rewritten as
\begin{equation}
\frac{J^{2}}{2r_{+}^{3}}dr_{+}+\frac{f^{\prime }(r_{+})}{4\pi
}d(4\pi r_{+})- \frac{2r_{+}dr_{+}}{\ell ^{2}}=-P_{r}(2\pi
r_{+}dr_{+}). \label{12}
\end{equation}
Now using equation $\left( \ref{9}\right) $ and value of
$J=2\Omega r_{+}^{2} $ \ ($dJ=4\Omega r_{+}dr_{+}$) in the above
equation we have
\begin{equation}
4\Omega ^{2}r_{+}dr_{+}+\frac{f^{\prime }(r_{+})}{4\pi }d(4\pi
r_{+})+\frac{ \Phi }{2}dQ-dM=-P_{r}d(\pi r_{+}^{2}).  \label{13}
\end{equation}
The first two terms on the left hand side of equation $\left(
\ref{13} \right) $ are trivial to recognize as $\Omega dJ\ $and
$TdS$ respectively. As discussed in reference  \cite{JZ}, if we
similarly identify $M+\frac{\Phi }{2}Q$ as the total energy $E$
and $A$ as the area enclosed by the horizon, then equation $\left(
\ref{13}\right) $ can be written as
\begin{equation}
dE=TdS+\Omega dJ+\Phi dQ+P_{r}dA,  \label{14}
\end{equation}
which is identical to the first law of thermodynamics. Hence the
field equations near horizon of CR-BTZ black hole behave like a
thermal system satisfying the first law of thermodynamics.

\section{Conclusion}

In this paper, we have discussed the CR-BTZ black hole
particularly in context with its horizons. The presence of
logarithmic term in the lapse function makes it difficult to find
exact expressions for the horizons. However, by qualitative
analysis of the lapse function, we are able to obtain conditions
on $M$, $J$, and $Q$ for the existence of naked singularity,
extreme and usual CR-BTZ black hole cases. It is also shown that
the particular cases, (i) $J=0$ and (ii) $Q=0$, of our results
reduce to the earlier known results for charged and rotating BTZ
black holes respectively.

The expression for the heat capacity, $C_{J,Q}$, for the CR-BTZ
black hole is obtained. It is shown that the heat capacity is
positive definite for the usual CR-BTZ black hole and zero for the
extreme case.

As mentioned in the introduction, the field equations for
spherically symmetric geometry, in case of Einstein and
Lanczos-Lovelock theories of gravity can be expressed as
$TdS=dE+PdV$. For the Kerr-Newmann geometry, which is axially
symmetric, the field equations were shown to hold the identity
$TdS=dM-\Omega dJ+\Phi dQ$. Here we have further followed this
approach for the (2+1)-dimensional CR-BTZ black hole. It is shown
that near horizon the Einstein field equations can be expressed as
a first law of thermodynamics $dE=TdS+\Omega dJ+\Phi dQ+P_{r}dA$,
where $E=M+\frac{\Phi}{2}Q$ is the total energy of the black hole.
These thermal identities indicate intrinsic thermodynamic
properties of spacetime horizons.

\section{Acknowledgments}

M. Akbar would like to acknowledge the research facilities
provided by NUST Centre for Advanced Mathematics and Physics
during this work. Authors are also grateful to the referee for
his/her comments which have significantly improved quality of the
paper.

\end{document}